\documentclass[runningheads,a4paper]{llncs}
\usepackage{threeparttable}
\usepackage{amssymb}
\usepackage{graphicx}
\usepackage{color}
\usepackage{comment}
\usepackage{url} 
\usepackage{multirow}

\newenvironment{list1}{\begin{list}{$\bullet$}
{\topsep 0 pt \parsep 0 pt \partopsep 0 pt \itemsep 0
pt}}{\end{list}}

\usepackage{times}

\catcode`~=11 
\newcommand{\urltilde}{\kern -.15em\lower .7ex\hbox{~}\kern .04em}
\catcode`~=13 

\newcommand{\G}{\textbf{G}\xspace}
\newcommand{\F}{\textbf{F}\xspace}

\newcommand{\U}{\textbf{U}\xspace}
\newcommand{\sig}[1]{\textsf{#1}\xspace}

\newcommand{\X}{\textbf{\textsc{X}}}
\newcommand{\s}[1]{\textsf{{#1}}}
\newcommand{\A}{\mathcal{A}}
\newcommand{\game}{\mathcal{G}}
\usepackage[ruled,vlined]{algorithm2e}
\usepackage{xspace}
\newcommand{\comments}[1]{}
\newcommand{\keywords}[1]{\par\addvspace\baselineskip
\noindent\keywordname\enspace\ignorespaces#1}

\begin{document}

\mainmatter  

\title{\textsf{G4LTL-ST}: Automatic Generation of PLC Programs}
\titlerunning{\textsf{G4LTL-ST}: Automatic Generation of PLC Programs}

%
%
\author{
Chih-Hong Cheng\inst{1}
\and
Chung-Hao Huang\inst{2}
\and
Harald Ruess\inst{3}
\and
Stefan Stattelmann\inst{1}
}

\authorrunning{C.-H.~Cheng, C.-H.~Huang, H.~Ruess, and S.~Stattelmann}

\institute{
ABB Corporate Research, Ladenburg, Germany
\and
Department of Electrical Engineering, National Taiwan University, Taipei, Taiwan
\and
fortiss - An-Institut Technische Universit{\"a}t  M\"{u}nchen, M\"{u}nchen, Germany
}
\maketitle

\vspace{-5mm}

\begin{abstract}
\textsf{G4LTL-ST} automatically synthesizes  control code for
industrial Programmable Logic Controls (PLC) from timed behavioral specifications of  input-output signals.
These specifications are expressed in a linear temporal logic (LTL) extended with non-linear arithmetic constraints
and timing constraints on signals\@.
\textsf{G4LTL-ST} generates code in IEC 61131-3-compatible {\em Structured Text}, which is compiled
into executable code for a large  number of industrial field-level devices.
The synthesis algorithm of \textsf{G4LTL-ST} implements pseudo-Boolean abstraction of data constraints and the
compilation of timing constraints into LTL, together with  a counterstrategy-guided abstraction-refinement synthesis loop\@.
Since  temporal logic specifications are notoriously difficult to use in practice, \textsf{G4LTL-ST} supports
engineers in specifying realizable control problems by suggesting suitable restrictions
on the behavior of the control environment from failed synthesis attempts.


\end{abstract}

\vspace{-5mm}
\section{Overview}
\vspace{-2mm}

Programmable Logic Controllers (PLC) are ubiquitous in the manufacturing and processing industries for realizing
real-time controls with stringent dependability and safety requirements.
A PLC is designed to read digital and analog inputs from various sensors and other PLCs, execute a user-defined program,
and write the resulting  digital and analog output values to various output elements including hydraulic and pneumatic actuators or indication lamps\@.
The time it takes to complete such a scan cycle typically ranges in the milliseconds\@.

The languages defined in the IEC 61131-3 norm
are the industry standard for programming PLCs~\cite{IEC61131-3}.
Programming in these rather low-level languages can be very inefficient, and yields inflexible controls which are
difficult to maintain and arduous to port.
Moreover, industry is increasingly moving towards more flexible and modular
production systems, where the control software is required to adapt to frequent specification changes~\cite{acatech2013report}\@.

\begin{figure}[t]
1\,\,\,\,\,\,\,\,\,\,\,\, Input: $\sig{x}, \sig{y}\in [0, 4] \cap \mathbb{R}$,  $\sig{err}\in \mathbb{B}$,\,\, Output: $\sig{grant1}, \sig{grant2}, \sig{light}\in \mathbb{B}$,\,\, Period: $\sig{50ms}$\\
2\\
3\,\,\,\,\,\,\,\,\,\,\,\, $\G\, (\sig{x}+\sig{y}>3 \rightarrow \textbf{X}\, \sig{grant1})$ \\
4\,\,\,\,\,\,\,\,\,\,\,\, $\G\, (\sig{x}^2+\sig{y}^2<\frac{7}{2} \rightarrow \textbf{X}\, \sig{grant2})$\\
5\,\,\,\,\,\,\,\,\,\,\,\, $\G\, (\neg(\sig{grant1} \wedge \sig{grant2}))$\\
6\,\,\,\,\,\,\,\,\,\,\,\, $\G\, (\sig{err} \rightarrow \sig{10sec}(\sig{light})$)\\
7\,\,\,\,\,\,\,\,\,\,\,\, $\G\, ((\G \neg\sig{err}) \rightarrow (\F \G \neg\sig{light}$))

\caption{Linear temporal logic specification with arithmetic constraints and a timer.}
\label{fig.original.spec}
\vspace{-5mm}
\end{figure}

With this motivation in mind, we developed the synthesis engine~\textsf{G4LTL-ST} for
generating IEC 61131-3-compatible  Structured Text programs from behavioral specifications.
Specifications of industrial control problems are expressed in a suitable extension of linear temporal logic (LTL)~\cite{pnueli1977temporal}\@.
The well-known LTL operators \G, \F, \U, and \textbf{X} denote ``always'', ``eventually'', ``(strong) until'', and ``next''s relations
over linear execution traces\@.
In addition to vanilla LTL, specifications in~\textsf{G4LTL-ST} may also include
    \begin{itemize}
    \item non-linear arithmetic constraints for specifying non-linear constraints on real-valued inputs; 
    \item timing constraints based on timer constructs specified in IEC 61131-3\@.
    \end{itemize}
A timing constraint of the form \sig{10sec(light)}, for example, specifies that the \sig{light} signal is on for 10 seconds.
Moreover, the semantics of temporal specifications in \textsf{G4LTL-ST} is slightly different from the standard
semantics as used in model checking,
since the execution model of PLCs is based on the concept of~\emph{Mealy machines}\@.
Initial values for output signals are therefore undefined, and the synthesis engine of~\textsf{G4LTL-ST} assumes that the environment of the controller makes the first move by setting the inputs.

Consider, for example, the PLC specification in Figure~\ref{fig.original.spec} with a specified scan cycle time of \textsf{50ms} (line~1)\@.
The input variables  $\sig{x}, \sig{y}, \sig{err}$ store bounded input and sensor values, and
output values are available at the end of each scan cycle at $\sig{grant1}$, $\sig{grant2}$, and $\sig{light}$ (line~1)\@.
According to the specification in line~6, the  output $\sig{light}$ must be on for at least~10 seconds whenever an
error occurs, that is, input signal $\sig{err}$ is raised. Line~7 requires that if \textsf{err} no longer appears, then eventually the \sig{light} signal is always off\@.
The transition-style LTL specifications~3 and~4  in Figure~\ref{fig.original.spec} require setting $\sig{grant1}$ (resp.  $\sig{grant2}$)
to \sig{true} in the next cycle whenever the condition $\sig{x}+\sig{y}>3$ (resp. $\sig{x}^2+\sig{y}^2<\frac{7}{2}$)  holds\@.
Finally,  $\sig{grant1}$ and $\sig{grant2}$ are supposed to be mutually exclusive (line~5)\@.

The synthesis engine of~\textsf{G4LTL-ST} builds on top of traditional LTL synthesis techniques~\cite{pnueli1989synthesis,Jobstm06c,ScheweF07a,acacia12}
which view the synthesis problem as a game between the (sensor) environment and the controller\footnote{Appendex~\ref{sec.LTL.synthesis} provides the detailed formulation and implemented algorithm for LTL synthesis.}.  
The moves of the environment in these games are determined by setting the input variables, and the
controller reacts by setting output variables accordingly.
The controller wins if the resulting input-output traces satisfy the given specification.
Notably, arithmetic constraints and timers are viewed as theories and thus abstracted
into a pseudo-Boolean LTL formula. This enables \textsf{G4LTL-ST} to utilize CEGAR-like~\cite{clarke2000counterexample,nieuwenhuis2005abstract,henzinger03CEC} techniques for successively constraining the capabilities of the control environment\@.

Since specifications in linear temporal logic are often notoriously difficult to use in practice, \textsf{G4LTL-ST}
diagnoses unrealizable specifications and suggests additional~\emph{assumptions}
for making the controller synthesis problem realizable.  The key hypothesis underlying this approach is that this kind of
feedback is more useful for  the  engineer compared to, say,  counter strategies.
The assumption generation of~\textsf{G4LTL-ST}  uses built-in templates and heuristics
for estimating the importance and for ordering the generated assumptions accordingly.

Synthesis of control software, in particular, has been recognized as a key {\em Industrie 4.0} technology for realizing flexible and modular
controls (see, for example,~\cite{dke2013report}, RE-2 on page 44)\@.
The synthesis engine \textsf{G4LTL-ST} is planned to be an integral part of a  complete development tool chain towards
meeting  these challenges.
 \textsf{G4LTL-ST} is written in Java and is available
 (under the GPLv3 open source license) at
 \vspace{-1mm}
\begin{verbatim}        http://www.sourceforge.net/projects/g4ltl/files/beta\end{verbatim}
 \vspace{-1mm}

In the following we provide an overview of the main features of  \textsf{G4LTL-ST} including  Pseudo-Boolean abstractions of timing constraints, the abstraction-refinement synthesis loop underlying~\textsf{G4LTL-ST} and its implementation, and, finally, the template-based generation for suggesting new constraints of the behavior of the environment for making the control synthesis problem realizable.
These features of~\textsf{G4LTL-ST}  are usually only illustrated by means of examples, but the initiated reader
should be able to fill in missing technical details.

\begin{table}[t]
\begin{small}
\begin{center}
    \begin{tabular}{|l|l|}
    \hline
    {\bf Real-time specification pattern}                               &  {\bf Encoding in LTL} \\[0mm] \hline\hline
     Whenever \sig{a}, then  \sig{b} for $t$ seconds                    & $\G\, (\sig{a} \rightarrow (\sig{t1.start}\wedge
     \sig{b} \wedge \textbf{X} (\sig{b}\, \U \,\sig{t1.expire})))$                        \\ \hline
    Whenever \sig{a} continues for more  &  $(\sig{a} \leftrightarrow \sig{t1.start}) \wedge
                     \G  (\neg(\sig{a} \wedge \textbf{X}\, \sig{a})  \leftrightarrow  \textbf{X}\, \sig{t1.start})   $                  \\
          than $t$ seconds,  then \sig{b}            & $\wedge\,\G (\sig{t1.expire}  \rightarrow  \sig{b}) $                             \\ \hline
    Whenever \sig{a}, then \sig{b},            & $\G (\sig{a} \leftrightarrow \sig{t1.start}) \wedge
    \G (\neg(\sig{c} \wedge \textbf{X}\, \sig{c}) \leftrightarrow \textbf{X}\, \sig{t1.start})$\\
    until \sig{c} for more than $t$ seconds    & $\wedge\, \G \,(\sig{a} \rightarrow (\sig{b} \wedge \textbf{X}((\sig{b} \,\U\, \sig{t1.expire})) \vee \G \neg \sig{t1.expire}))$                               \\ \hline
    \end{tabular}
\end{center}
\end{small}
\label{table:translation}
\caption{Real-time specification patterns and their encodings.}
 \vspace{-5mm}
\end{table}

\vspace{-2mm}
\section{Timing Abstractions}~\label{sec:timers}
\vspace{-2mm}

The timing constraint in Figure~\ref{fig.original.spec} with its 10 seconds time-out may be encoded  in LTL by associating
each discrete step with a 50ms time delay. Notice, however, that up to 200 consecutive \textbf{X} operators are needed for
encoding this simple example.

Instead we propose a more efficient translation, based on standard IEC 61131-3 timing constructs, for realizing timing specifications.
Consider, for example, the timed specification $\G\, (\sig{err} \rightarrow \sig{10sec}(\sig{light}))$\@.
In a first step, fresh variables $\sig{t1.start}$ and $\sig{t1.expire}$ are introduced, where  $\sig{t1}$ is a \emph{timer variable} of type \sig{TON} in IEC 61131-3\@.
The additional output variable $\sig{t1.start}$ starts the timer $\sig{t1}$, and the additional input variable $\sig{t1.expire}$ receives  a time-out signal from  $\sig{t1}$ ten seconds
after this timer has been started\@.
Now, the timing specification $\G\, (\sig{err} \rightarrow \sig{10sec}(\sig{light}))$ is rewritten as an LTL specification for a function block in the context of a timer.
\[
\G\, (\sig{t1.start} \rightarrow \textbf{X}\, \F\, \sig{t1.expire}) \rightarrow
\G\, (\sig{err} \rightarrow (\sig{t1.start} \wedge \sig{light} \wedge \textbf{X} (\sig{light}\, \U \,\sig{t1.expire}))
\]
The antecedent formula ensures that the \textsf{expire} signal is eventually provided by the timing block of the environment.
Since no provision is being made that there is a time-out exactly after 10 seconds, however, the precise expected behavior
of the time-out environment is over-approximated.


It is straightforward to generate PLC code using timing function blocks from winning strategies of the controller.
Whenever  $\sig{t1.start}$ is set to $\sig{true}$ the instruction \begin{small}\textsf{t1(IN:=0, PT:=TIME\#10s)}\end{small} is
generated for starting the timer \sig{t1}. Instructions that set \sig{t1.start} to \sig{false} is ignored based on the underlying semantics of timers. Finally, time-out signals \sig{t1.expire} are simply replaced with the variable $\sig{t1.Q}$ of the IEC 61131-3 timing construct.

In Table~\ref{table:translation} we describe some frequently encountered specification patterns and their translations using  IEC 61131-3-like
timing constructs. Each of these patterns requires the introduction of a fresh timer variable \sig{t1} together with the assumption
$\G\, (\sig{t1.start} \rightarrow \textbf{X}\, \F\, \sig{t1.expire})$ on the environment providing time-outs\@.
These specification patterns, however, are not part of the \textsf{G4LTL-ST} input language, since there is
no special support in the synthesis engine for these language constructs,  and \textsf{G4LTL-ST} is intended to
be used in integrated development frameworks, which usually come with their own specification languages.

\vspace{-2mm}
\section{Abstraction-Refinement Synthesis Loop}\label{subsec:numerical}
\vspace{-2mm}

\begin{figure}[t]
\centering
\includegraphics[width=0.6\columnwidth]{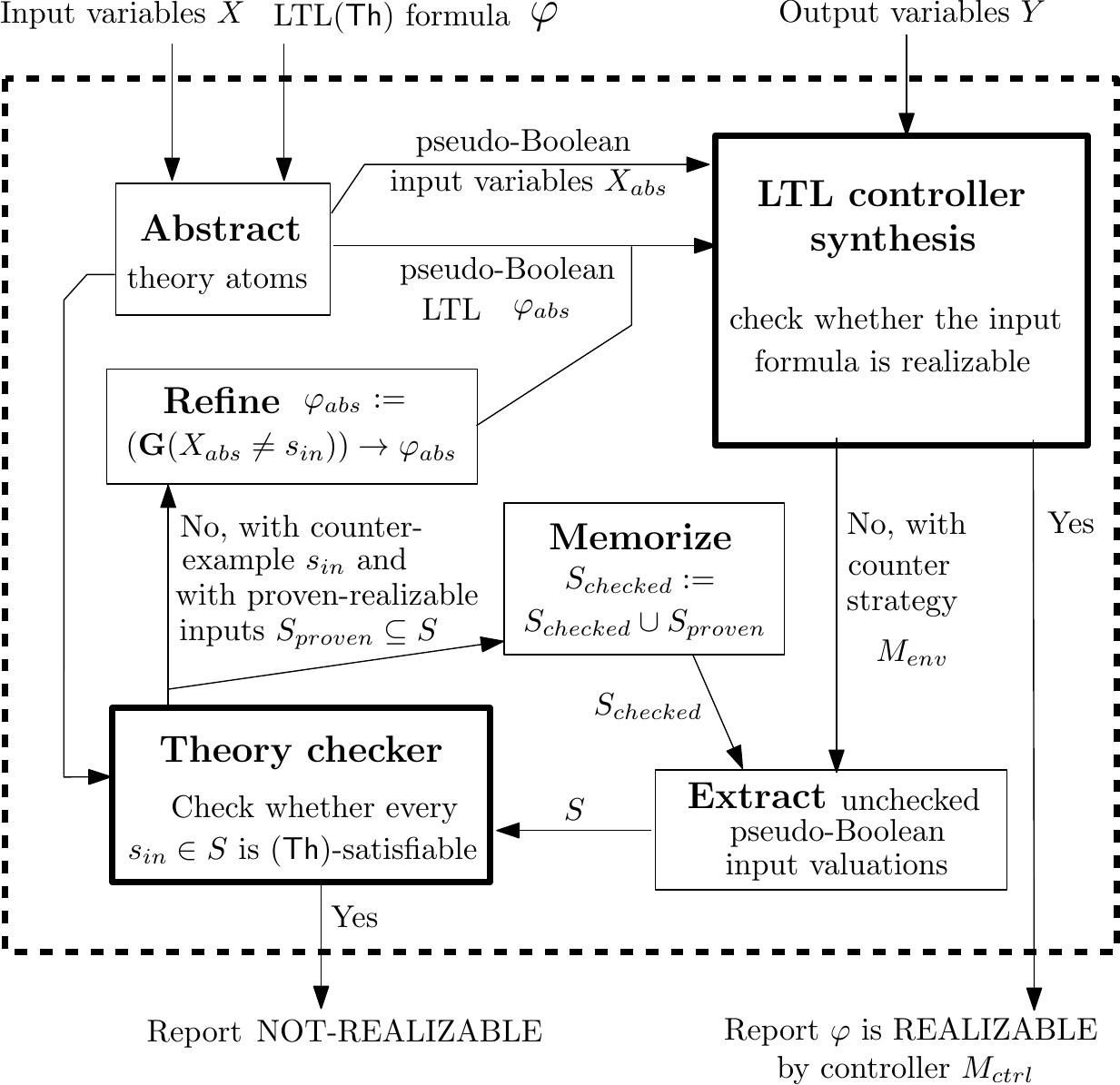}
\caption{Abstraction-refinement synthesis loop.}
\label{fig:CEGAR}
\vspace{-5mm}
\end{figure}

The input to the synthesis engine of \textsf{G4LTL-ST} are  LTL formulas with non-linear arithmetic constraints with
bounded real (or rational) variables, and the workflow of this engine is depicted in Figure~\ref{fig:CEGAR}\@.
Notice, however, that the abstraction-refinement loop in Figure~\ref{fig:CEGAR} is more general in that it works for any decidable theory {\textsf Th}\@.

In a preliminary step \textsf{Abstract} simply replaces arithmetic constraints on the inputs with fresh Boolean input variables.
The resulting specification therefore is (like the timer abstraction in Section~\ref{sec:timers}) an over-approximation of the behavior of the environment.
In our running example in Figure~\ref{table:translation} (ignoring line 6, 7), \textsf{Abstract} creates two fresh Boolean
variables, say $\sig{req1}$ and $\sig{req2}$,  for the two input constraints $\sig{x}+\sig{y}>3$ and $\sig{x}^2+\sig{y}^2<\frac{7}{2}$
to obtain the pseudo-Boolean specification
\begin{equation}\label{eq.abstract.spec}
 \G (\sig{req1} \rightarrow \textbf{X}\, \sig{grant1})
 \wedge \G (\sig{req2} \rightarrow \textbf{X}\, \sig{grant2}) \wedge
 \G (\neg(\sig{grant1} \wedge \sig{grant2}))
\end{equation}
Clearly, this pseudo-Boolean specification with input variables $\sig{req1}$ and $\sig{req2}$
over-approximates the behavior of the environment, since it does not account for inter-relationships of the arithmetic input constraints.

In the next step,  \textsf{LTL controller synthesis} checks whether or not the pseudo-Boolean LTL formula generated by  \textsf{Abstract} is realizable.
If the engine is able to realize a winning strategy for the control, say $M_{ctrl}$,  then a controller is  synthesized from this strategy.
Otherwise, a candidate counter-strategy, say $M_{env}$,  for defeating the controller's purpose is generated.

The pseudo-Boolean specification~(\ref{eq.abstract.spec}), for example, is unrealizable.
A candidate counter-strategy for the environment is given by only using the input $(\sig{true}, \sig{true})$,
since, in violation of the mutual exclusion condition~(\ref{eq.abstract.spec}), the controller is forced to subsequently set both \sig{grant1} and \sig{grant2} \@.

The \textsf{Extract} module extracts candidate counter-strategies with fewer
pseudo-Boolean input valuations (via a greedy-based method) whose validity are not proven at the theory level.
Consequently, the \textsf{Extract} module generates a candidate counter-strategy that only uses $(\sig{req1}, \sig{req2})=(\sig{true}, \sig{true})$
and the input valuations $S=\{(\sig{true}, \sig{true})\}$  are passed to the {\textsf Theory Checker\@.

A candidate counter-strategy is a genuine counter-strategy only if all pseudo-Boolean input patterns are satisfiable at the theory level; in
these cases the environment wins and {\textsf Theory Checker} reports the un-realizability of the control problem.
In our running example, however,  the input $(\sig{true}, \sig{true})$ is not satisfiable at the theory level,
since the conjunction of the input constraints $\sig{x}+\sig{y}>3$ and $\sig{x}^2+\sig{y}^2<\frac{7}{2}$ is unsatisfiable for $\sig{x}, \sig{y} \in [0, 4]$\@.
\textsf{G4LTL-ST} uses the \textsf{JBernstein}~\cite{jbernstein} verification engine for discharging
quantifier-free verification conditions  involving non-linear real arithmetic\@.
In order to avoid repeated processing at the theory level, all satisfiable inputs are memorized.

Unsatisfiable input combinations $s_{in}$ are excluded by \textsf{Refine}\@.
In our running example, the formula $\G (\neg(\sig{req1} \wedge \sig{req2}))$ is added as a new assumption on the environment,
since  the input pair $(\sig{true}, \sig{true})$ has been shown to be unsatisfiable.
\begin{equation}\label{eq.abstract.spec.refined}
 \G (\neg(\sig{req1} \wedge  \sig{req2})) \rightarrow  \mbox{(\ref{eq.abstract.spec})}
\end{equation}
In this way, \textsf{Refine} successively refines the over-approximation of the behavior of the environment.
Running  the {\textsf LTL synthesis engine} on the refined specification~\ref{eq.abstract.spec.refined} yields a controller:
if one of \sig{req1} ($\sig{x}+\sig{y}>3$) and \sig{req2} ($\sig{x}^2+\sig{y}^2<\frac{7}{2}$) holds, the controller may grant the
corresponding  client in the next round, since \sig{req1} and \sig{req2} do not  hold simultaneously.

\vspace{-2mm}
\paragraph{Refinement of Timer Environments.}
The refinement of over-approximations of environmental behavior also works for the abstracted timer environments.
Recall from Section~\ref{sec:timers} that the initial abstraction is given by $\G\, (\sig{t1.start} \rightarrow \textbf{X}\, \F\, \sig{t1.expire})$\@.
Assuming, for example, that \sig{t1.expire} appears two iterations after \sig{t1.start} in a candidate counter-strategy, one might strengthen this initial
assumption with
$\G\, (\sig{t1.start} \rightarrow ((\textbf{X} \neg\sig{t1.expire}) \wedge (\textbf{X}\textbf{X} \neg\sig{t1.expire}) \wedge (\textbf{X}\textbf{X}\textbf{X}\, \F\, \sig{t1.expire})))\mbox{.}$

\vspace{-2mm}
\paragraph{Synthesized Code.}
The synthesized PLC for the complete control specification in Figure~\ref{fig.original.spec} is listed in the Appendix~\ref{app.synthesized.code}\@.
This synthesized function block can readily be passed to industry-standard PLC development tools for connecting function blocks with concrete field device signals inside the main program to demonstrate desired behavior.
Notice that the synthesized code in Appendix~\ref{app.synthesized.code} is separated into two independent state machines.
This separation is based on a simple analysis of \textsf{G4LTL-ST} for partitioning the specification into blocks with mutually independent output variables. In particular, in Figure~\ref{fig.original.spec}, the formulas in line~3 to 5 do not consider \sig{err} and \sig{light}, and the formulas in lines~6 and~7
do not consider \sig{x}, \sig{y}, \sig{grant1}, \sig{grant2}\@. Therefore, code for these two blocks can be synthesized independently.

\vspace{-2mm}
\paragraph{Constraints over input and output variables.}
Even though the current implementation of \textsf{G4LTL-ST} is restricted to specifications with arithmetic constraints on inputs only,
the abstraction-refinement synthesis loop in Figure~\ref{fig:CEGAR} works more generally for arithmetic constraints over input and
output variables.
Consider, for example, the specification $\G(\sig{x} > \sig{y} \rightarrow \textbf{X} (\sig{z} > \sig{x}))$ with
 input variables $\sig{x}, \sig{y} \in [1, 2]\cap \mathbb{R}$ and output variable $\sig{z} \in [0, 5]\cap \mathbb{R}$\@.
Abstraction yields a pseudo-Boolean specification  $\G (\sig{in} \rightarrow \textbf{X} \sig{out})$
with  \sig{in}, \sig{out} fresh input variables  for the constraints $\sig{x} > \sig{y}$ and  $\sig{z} > \sig{x}$, respectively.
Now, pseudo-Boolean LTL synthesis generates a candidate winning strategy $M_{ctrl}$ for the controller, which simply sets the
output $\sig{out}$ to be always  $\sig{true}$\@.
The candidate controller $M_{ctrl}$ is realizable if every pseudo-Boolean output assignment  of $M_{ctrl}$ is indeed satisfiable on the theory level. This condition amounts to demonstrating validity of the quantified formula
        $(\forall \sig{x} \in [1, 2]\cap \mathbb{R})\, (\exists \sig{z}\in [0,5]\cap \mathbb{R})\, \sig{z}>\sig{x}\mbox{.}$
Using the witness, say,  $3$ for the existentially quantified output variable $z$, a winning strategy
for the  controller is to always set the output $z$ to $3$, and the control synthesis problem therefore is realizable.

Otherwise, the candidate controller strategy is not realizable at the theory level, and,
for pseudo-Boolean outputs, refinement due to un-realizability of the control synthesis problem
is achieved by adding new
constraints as \emph{guarantees} to  the pseudo-Boolean specification.
For example the constraint $\G (\neg(\sig{grant1} \wedge \sig{grant2}))$ is added to the pseudo-Boolean
specification, if pseudo-Boolean outputs $\sig{grant1}$ and $\sig{grant2}$ are mutually exclusive at the theory
level.

In this way, the abstraction-refinement synthesis loop in Figure~\ref{fig:CEGAR} may handle arbitrary theory constraints
on input and output  variables as long as corresponding verification conditions in a first-order theory with one
quantifier-alternation can be decided. The implementation of \textsf{G4LTL-ST} could easily be extended in this
direction by using, for examples the verification procedure for the exists-forall fragment of non-linear arithmetic as
described in ~\cite{efsmt}\@. So far we have not yet encountered the need for this extensions, since
the PLC case studies currently available to us are restricted to Boolean outputs.

\begin{table}
\label{fig.eval}
\begin{scriptsize}
\centering
\begin{tabular}{|c||c|c|c|c|}
\hline
 \parbox[t]{1.3cm}{\centering{\bf \# Example}\\\centering{\bf (synthesis)}}  & \parbox[t]{1.6cm}{\centering{\bf Timer(T)/}\\\centering{\bf Data(d)}} & \parbox[t]{2cm}{\centering{\bf lines of spec}} &\parbox[t]{1.3cm}{\centering{\bf Synthesis}\\\centering{\bf Time}} & \parbox[t]{1.2cm}{\centering{\bf Lines of}\\\centering{\bf ST}}   \\ \hline\hline
Ex1                   &            T, D        &       9            &           1.598s (comp)            &          110         \\
Ex2                    &           T          &       13            &          0.691s             &      148             \\
Ex3                    &           T          &       9            &          0.303s             &       80             \\
Ex4                    &           T          &       13            &         21s              &       1374            \\
Ex5                    &           T          &       11            &        0.678s (comp)           &        210               \\
Ex6                    &           -          &       7            &      0.446s                 &     41              \\
Ex7                    &           D          &        8           &          17s             &        43           \\
Ex8                    &           T          &        8           &         0.397s (comp)             &       653            \\
Ex9               &   abstract D,T        &    3 + model ($<200$ loc)            &       1.55s                &        550           \\
Ex10             &    abstract D,T    &   3 + model ($<200$ loc)   &        3.344s               &   229   \\
Ex11             &    abstract D,T    &   3 + model ($<200$ loc)   &        0.075s               &   105  \\
\hline
\end{tabular}
\centering
\vspace{2mm}
\centering
\begin{tabular}{|c||c|c|}
\hline
 \parbox[t]{1.3cm}{\centering{\bf \# Example}\\\centering{\bf (Assup. gen)}}  & \parbox[t]{1.2cm}{\centering{\bf \# Learned}\\\centering{\bf Assump.}} & \parbox[t]{1.2cm}{\centering{\bf Time of}\\ \centering{\bf Learning}}  \\ \hline\hline
Ex1                    &        1      &   0.127s      \\
Ex2                    &        1      &  0.452s      \\
Ex3                    &        1      &  3.486s      \\
Ex4                    &        4      &  22s (DFS)     \\
Ex5                    &       1       &   2.107s     \\
Ex6                    &       1       &   1.046s     \\
Ex7                    &       1       &   0.154     \\
Ex8                    &       1       &   2.877     \\
Ex9                   &       1       &   8.318     \\
\hline
\end{tabular}
\centering

\label{table:experiment}
\end{scriptsize}
\vspace{3mm}
\caption{Experimental result based on the predefined unroll depth (3) of \textsf{G4LTL-ST}. Execution time annotated with ``(comp)'' denotes  that the value is reported by the compositional synthesis engine.}
\end{table}

\vspace{-2mm}
\section{Assumption Generation}~\label{ref.assumption.generation}
An unrealizable control synthesis problem can often be made realizable by restricting the capabilities of the input environment
in a suitable way. In our case studies from the manufacturing domain, for example, suitable restrictions on the arrival rate of workpieces were often helpful. \textsf{G4LTL-ST} supports the generation of these assumptions from a set of given templates.
For example, instantiations of the template  $\G (\sig{?a} \rightarrow (\textbf{X} (\neg \sig{?a} \,\U \,\sig{?b})))$,
where $\sig{?a}$ and  $\sig{?b}$ are meta-variables for inputs, disallows successive arrivals of  an input signal $\sig{?a}$\@.
For a pre-specified set of templates, \textsf{G4LTL-ST} performs a heuristic match of the meta-variables with
input variables by analyzing possible ways of the environment to defeat  the control specification.


The underlying LTL synthesis engine performs bounded unroll~\cite{ScheweF07a} of the negated property to safety games.
Therefore, whenever the controller can not win the safety game, there exists an environment strategy which can be expanded as a finite tree, whose leaves are matched with the risk states of the game. Then,  the following three steps are performed successively:
\begin{list1}
    \item \emph{Extract} a longest path from the source to the leaf. Intuitively, this path represents a scenario where the controller
             endeavors to resist losing the game (without intentionally losing the game). For example, assume for such a longest path,
             that the environment uses $(\sig{a})(\neg\sig{a})(\neg\sig{a})(\neg\sig{a})$ to win the safety game.
    \item \emph{Generalize} the longest path. Select from the set of templates one candidate which can {\em fit} the path
             in terms of  generalization. For example, the path above may be generalized as $\F\G\neg\sig{a}$\@.
             For every such template, the current implementation of \textsf{G4LTL-ST} defines a unique generalization function.
    \item \emph{Resynthesize} the controller based on the newly introduced template.  For example, given $\phi$ as the original
             specification, the new specification will be $(\neg\F\G\neg\sig{a}) \rightarrow \phi$,
              which is equivalent to $(\G\F\sig{a}) \rightarrow \phi$\@.
             Therefore, the path is generalized as an assumption stating that $\sig{a}$ should appear infinitely often.
\end{list1}
If this process fails to synthesize a controller, then new assumptions are added to further constrain the environment behavior.
When the number of total assumptions reaches a pre-defined threshold but no controller is generated, the engine stops and reports its inability to decide the given controller synthesis problem.

\vspace{-2mm}
\section{Outlook}
\vspace{-2mm}
The synthesis engine of \textsf{G4LTL-ST} has been evaluated on a number of simple automation examples extracted
both from public sources and from ABB internal projects\@.
The evaluation results in Table~2 demonstrate that,
despite the underlying complexity of the LTL synthesis,~\textsf{G4LTL-ST}
can still provide a practical alternative to the prevailing low-level encodings of PLC programs\footnote{Short descriptions of the case studies have been added to the appendix.}, whose block size are (commonly) within 1000 LOC. This is due to the fact that many modules are decomposed to only process a small amount of I/Os.
For small sized I/Os, the abstraction of timers and data in~\textsf{G4LTL-ST} together with counter-strategy-based lazy refinement are particularly effective in fighting the state explosion problem, since unnecessary unrolling (for timing) and bit-blasting (for data) are avoided\@. Data analysis is also effective when no precise (or imprecise) environment model is provided, as is commonly the case in industrial automation scenarios.

Mechanisms such as assumption generation are essential for the wide-spread deployment of~\textsf{G4LTL-ST} in industry,
since they provide feedback to the designer in the language of the problem domain.
Extensive field tests, however, are needed for calibrating assumption generation in practice.
Moreover, a targeted front-end language for high-level temporal specification of typical control problems for (networks of) PLCs
needs to be developed~\cite{ljungkrantz2010PLC}\@.


\appendix


\newpage
\vspace{-3mm}

\section{Synthesized Function Block}  \label{app.synthesized.code}
\vspace{-3mm}


\noindent
Structured Text generated by \textsf{G4LTL-ST} for the synthesis control problem in Figure~\ref{fig.original.spec}\@.

\begin{scriptsize}
\begin{verbatim}
FUNCTION_BLOCK FB_G4LTL
VAR_INPUT
     x: REAL; y: REAL;
     error: BOOL;
END_VAR

VAR_OUTPUT
    grant1: BOOL; grant2: BOOL;
    light: BOOL;
END_VAR

VAR
   cstate1 : INT := 0;   cstate2 : INT := 0;
   p0 : BOOL; p1 : BOOL;
   t1: TON;
END_VAR

VAR CONST  T1_VALUE	: TIME := TIME#10s;   END_VAR

p0 := (x)+(y)>3;  p1 := (x*x)+(y*y)<3.5;     (* Update the predicate based on sensor values *)

CASE cstate1 OF   (* State machines *)
  0:  IF ((p0 = TRUE) AND  (p1 = FALSE)) THEN cstate1 := 9; grant1 := TRUE; grant2 := FALSE;
       ELSIF ((p0 = FALSE) AND  (p1 = FALSE)) THEN cstate1 := 0; grant1 := TRUE; grant2 := FALSE;
       ELSIF ((p0 = FALSE) AND  (p1 = TRUE)) THEN cstate1 := 7; grant1 := TRUE; grant2 := FALSE;
       ELSIF ((p0 = TRUE) AND  (p1 = TRUE)) THEN cstate1 := 11; grant1 := TRUE; grant2 := FALSE;
       END_IF;
  7:  IF ((p0 = TRUE) AND  (p1 = FALSE)) THEN cstate1 := 9; grant1 := FALSE; grant2 := TRUE;
       ELSIF ((p0 = FALSE) AND  (p1 = FALSE)) THEN cstate1 := 0; grant1 := FALSE; grant2 := TRUE;
       ELSIF ((p0 = FALSE) AND  (p1 = TRUE)) THEN cstate1 := 7; grant1 := FALSE; grant2 := TRUE;
       ELSIF ((p0 = TRUE) AND  (p1 = TRUE)) THEN cstate1 := 11; grant1 := TRUE; grant2 := FALSE;
       END_IF;
  9:  IF ((p0 = TRUE) AND  (p1 = FALSE)) THEN cstate1 := 9; grant1 := TRUE; grant2 := FALSE;
       ELSIF ((p0 = FALSE) AND  (p1 = FALSE)) THEN cstate1 := 0; grant1 := TRUE; grant2 := FALSE;
       ELSIF ((p0 = FALSE) AND  (p1 = TRUE)) THEN cstate1 := 7; grant1 := TRUE; grant2 := FALSE;
       ELSIF ((p0 = TRUE) AND  (p1 = TRUE)) THEN cstate1 := 11; grant1 := TRUE; grant2 := FALSE;
       END_IF;
 11: IF (( true )) THEN cstate1 := 11; grant1 := TRUE; grant2 := FALSE; END_IF;
END_CASE;

CASE cstate2 OF
   0: IF ((error = TRUE) AND  (TRUE)) THEN cstate2 := 12; light := TRUE; t1(IN:=0, PT:=T1_VALUE);
       ELSIF ((error = FALSE) AND  (TRUE)) THEN cstate2 := 6; light := FALSE;
       END_IF;
 43: IF ((error = TRUE) AND  (TRUE)) THEN cstate2 := 12; light := TRUE; t1(IN:=0, PT:=T1_VALUE);
       ELSIF ((error = FALSE) AND  (TRUE)) THEN cstate2 := 43; light := FALSE;
       END_IF;
   6: IF ((error = TRUE) AND  (TRUE)) THEN cstate2 := 12; light := TRUE; t1(IN:=0, PT:=T1_VALUE);
       ELSIF ((error = FALSE) AND  (TRUE)) THEN cstate2 := 6; light := FALSE;
       END_IF;
396: IF ((error = TRUE) AND  (TRUE)) THEN cstate2 := 12; light := TRUE; t1(IN:=0, PT:=T1_VALUE);
       ELSIF ((error = FALSE) AND  (t1.Q = FALSE)) THEN cstate2 := 396; light := TRUE;
       ELSIF ((error = FALSE) AND  (t1.Q = TRUE)) THEN cstate2 := 43; light := FALSE;
       END_IF;
 81: IF ((error = TRUE) AND  ( TRUE )) THEN cstate2 := 12; light := TRUE; t1(IN:=0, PT:=T1_VALUE);
       ELSIF ((error = FALSE) AND  (t1.Q = FALSE)) THEN cstate2 := 396; light := TRUE;
       ELSIF ((error = FALSE) AND  (t1.Q = TRUE)) THEN cstate2 := 43; light := FALSE;
       END_IF;
12: IF ((error = TRUE) AND  (TRUE)) THEN cstate2 := 12; light := TRUE; t1(IN:=0, PT:=T1_VALUE);
      ELSIF ((error = FALSE) AND  (t1.Q = FALSE)) THEN cstate2 := 81; light := TRUE;
      ELSIF ((error = FALSE) AND  (t1.Q = TRUE)) THEN cstate2 := 6; light := FALSE;
      END_IF;
END_CASE;
END_FUNCTION_BLOCK
\end{verbatim}
\end{scriptsize}


\section{Evaluation}  \label{app.evaluation}


In this section, we outline specifications for each problem under analysis.
\begin{itemize}
    \item \textbf{Ex1} Original example listed in Figure~\ref{fig.original.spec}.
    \item \textbf{Ex2} Simplified specification for \emph{equivalence with discrepancy time monitoring}: The discrepancy time is the maximum period during which both inputs may have different states without the function block detecting an error. The block should signal error when two input valuation differ in their values for more than time $T$.
    \item \textbf{Ex3} Simplified specification for \emph{safety request}: It monitors the response time between the safety
function request (\textsf{s\_opmode} set to \textsf{FALSE}) and the actuator acknowledgment (\textsf{s\_ack} switches to \textsf{TRUE}). Whenever the time exceeds the user-specified threshold $T$, signal \textsf{error} and propose again \textsf{request} to the actuator.
    \item \textbf{Ex4} Simplified specification for \emph{sequential muting}. Sequential muting is a standard mechanism to monitor the entry of workpieces into critical region using 4 sensors; it is used when transporting the material into the danger zone without causing the machine to stop. The mechanism should signal the entry and leave of objects, and if the object does not leave the critical region within a user-specified time interval $T$, error should be reported.
    \item \textbf{Ex5} Control for conveyor belt. Transfer an object when light barrier detects it. If another object appears before the current output leaves, signal light and make alarm sound for $T$ seconds. Resume when the \textsf{reset} button is pressed. Also retain silence when no object arrives.
    \item \textbf{Ex6} Example specification from the paper ``\emph{Formal Specification and Verification of Industrial Control Logic Components}''~\cite{ljungkrantz2010PLC}.
    \item \textbf{Ex7} Control synthesis for simple \emph{nonlinear systems}, where the underlying dynamics is captured by a predicate transition system, which can be considered as the result of timed relational abstraction~\cite{zutshi2012timed}. To successfully control the system, one needs to perform numerical reasoning also on the theory level, as the specification and the environment model use different predicates.
    \item \textbf{Ex8} Low-level \emph{device function synthesis}. Our previous work \textsf{MGSyn}~\cite{mgsyn} synthesizes high-level orchestration code (in C++) to control each cell via OPC DA communication\footnote{\url{http://en.wikipedia.org/wiki/OPC_Data_Access}}. Nevertheless, it assumes that each station (cell) should provide a set of pre-specified functions (as a library) to control low-level devices. As an experiment, we use \textsf{G4LTL-ST} to synthesize some these basic functions for the Pick\&Place module of the FESTO modular production system\footnote{http://didacticonline.de/int-en/learning-systems/mps-the-modular-production-system/stations/pick-place-station-small-is-beautiful.htm}.
        The combination of \textsf{MGSyn} and \textsf{G4LTL-ST} therefore completes the whole picture of automatic synthesis.

    \item \textbf{Ex9} The \emph{DC-DC boost converter} circuit example~\cite{senesky2003hybrid}. The goal is to synthesize a PLC controller which opens or closes switches appropriately such that the behavior of the underlying linear hybrid system is always constrained under a certain threshold. The environment model (as a transition system) is generated from timed relational abstraction~\cite{zutshi2012timed} extended with predicates\footnote{\textbf{NOTICE:} For this model, a special input format is used to read the result of relational abstraction and product it with the translated game, rather than viewing them as environmental assumptions (when doing so, the environmental assumption contains around 200 lines) and bring all assumptions into the process of generating B\"{u}chi automaton and perform bounded unroll.}. Two additional models are provided.
    \item \textbf{Ex10} Control of inverted pendulum with a physical model processed under timed-relational abstraction.
    \item \textbf{Ex11} Control of track system with a physical model processed under timed-relational abstraction.
    \item \textbf{Ex12} \textsf{ABB} Corporate Research in-house example (wind turbine yaw controller; details restricted).
    \item \textbf{Ex13} \textsf{ABB} Corporate Research in-house example (tide gate controller; details restricted).

    \item \textbf{AS} \emph{Assumption generation}. We use commonly seen examples (e.g., from Lily~\cite{Jobstm06c}) to demonstrate the use of assumption generation.  \textsf{G4LTL-ST} fails to synthesize controllers for these examples under the unroll depth 3 (some problems are realizable when increasing the unroll depth), but under this feature \textsf{G4LTL-ST} simple continues the synthesis process and discovers appropriate assumptions.
\end{itemize}

\section{LTL Synthesis Engine}\label{sec.LTL.synthesis}

In this section, we formulate required definitions the implemented algorithms for LTL synthesis. 

\paragraph{Linear Temporal Logic (LTL)} Given two disjoint set of atomic propositions $V_{in}= \{\textsf{i}_1, \ldots, \textsf{i}_m\}$ and $V_{out}= \{\textsf{o}_1, \ldots, \textsf{o}_n\}$,
the LTL  formulas over $V_{in}\cup V_{out}$ are defined inductively:
\begin{list1}
\item  $v \in V_{in}\cup V_{out}$ is an LTL formula.
\item  If $\phi_1, \phi_2$ are LTL-formulas, then so are $\neg \phi_1$, $\neg \phi_2$, $\phi_1 \vee \phi_2$, $\phi_1 \wedge \phi_2$, $\phi_1 \rightarrow \phi_2$.
\item  If $\phi_1, \phi_2$ are LTL-formulas, then so are  $\G \phi_1$, $\F \phi_1$, $\X \phi_1$, $\phi_1 \U \phi_2$.
\end{list1}
Given an $\omega$-word $\sigma \in \omega \rightarrow \mathbf{2}^{V_{in}\cup V_{out}}$ and a natural number $i \in \omega$,
the semantics of LTL is defined as follows.
\begin{list1}
\item $\sigma, i \vDash v$ iff $v \in \sigma(i)$.
\item $\sigma, i \vDash \neg \phi$ iff not $\sigma, i \vDash \phi$. Similar definitions follow for other logic operators $\vee, \wedge$, and $\rightarrow$. We use interchangeably $\neg$ and $\s{!}$.
\item $\sigma, i \vDash \G \phi$ iff for all $j\geq i$: $\sigma, j \vDash \phi$.
\item $\sigma, i \vDash \F \phi$ iff for some $j\geq i$: $\sigma, j \vDash \phi$.
\item $\sigma, i \vDash \X \phi$ iff $\sigma, i+1 \vDash \phi$.
\item $\sigma, i \vDash \phi_1 \U \phi_2$ iff for some $j\geq i$: $\sigma, j \vDash \phi_2$, and for all $k= i, \ldots, j-1$: $\sigma, k \vDash \phi_1$.
\end{list1}

\paragraph{$\omega$-automaton} A \emph{nondeterministic  B\"{u}chi word (NBW)} automaton over $\Sigma = \mathbf{2}^{V_{in}\cup V_{out}}$ is a finite automaton $\A = (Q, \Sigma, q_0, \Delta, F)$, where $Q$ is the set of states, $q_0$ is the initial state, $F\subseteq Q$ is the set of final states, and  $\Delta \subseteq Q\times \Sigma\times Q$ is the transition relation. A \emph{run} on an input word $\sigma = \sigma_1 \sigma_2 \ldots,$ where $\sigma \in \Sigma^{\omega}$, is a sequence $\varrho = \varrho_0 \varrho_1\ldots$ over states in $\A$ such that (1) $\varrho_0 = q_0$ and (2) for all $i\geq 0$: $(\varrho_i, \sigma_i, \varrho_{i+1}) \in \Delta$. $\A$ accepts $\sigma$ if there \emph{exists} a run $\varrho$ such that  $\varrho_i \in F$ for infinitely many $i$. A \emph{universal co-B\"{u}chi word (UCW)} automaton follows the above notation, but only accepts $\sigma$ if for \emph{every} run $\varrho$, $\varrho_i \in F$ for only finitely many $i$.

\paragraph{From LTL to $\omega$-automaton} It is known that for every LTL formula $\phi$, there exists an equivalent B\"{u}chi automaton representation $\mathcal{A}_{\phi}$, such that a word $\sigma \in \Sigma^{\omega}$ satisfies $\phi$ iff there exists an accepting run in $\mathcal{A}_{\phi}$ on $\sigma$ (algorithms can be found in, e.g.,~\cite{GaOd01,Giannakopoulou02fromstates}). Figure~\ref{fig:Example.LTL} shows some simple LTL formulas and the corresponding B\"{u}chi automata. Final states are with double circles. For the ease of understanding, in the graphical representation we fold many edges having the same source and destination into one, and represent these edges as a logic formula on the edge label. E.g., $\s{req}$ in Figure~\ref{fig:Example.LTL} actually represents the set $\{\langle\s{req,grant}\rangle, \langle\s{req,!grant}\rangle\}$.

\paragraph{Edge-labeled game} A labeled-game arena has the form $\game = (Q, Q_0, E, q_0, \Sigma_{in}, \Sigma_{out}, L)$, where $Q$ is the set of locations, $Q_0 \subset Q$ is the set of \emph{control} locations,  $E \subseteq Q\times Q$ is the transition relation, $q_0$ is the initial location. $L $ is a labeling function mapping from each transition with source in $Q_0$ to $\Sigma_{out}$, and each transition with source in $Q\setminus Q_0$ to $\Sigma_{in}$. Such a graph is partitioned into two areas $Q_0$ and $Q_1 = Q\setminus Q_0$ (\emph{environment} locations). A \emph{play} is a sequence $\rho = r_0 r_1 \ldots$ with $(r_i, r_{i+1}) \in E$ which is built up based on player selections: if $r_i \in Q_0$ then control (player-0) decides the next location, and if $r_i \in Q_1$ then environment (player-1) decides the next location.

A \emph{strategy} for control is a function $f: Q^{+}\rightarrow Q$ which allocates a play prefix $r_0 \ldots r_k$, $r_k \in Q_0$ a set $Q'\subseteq Q$  of vertices such that $(r_k, r_{k+1}) \in E$, for all $r_{k+1} \in Q'$. A strategy $f$ is winning for control at location $q_0$, if any play $\rho$ starting from $q_0$ following $f$ satisfies the \emph{winning condition}.
In this paper we only use safety condition: 
\begin{itemize}
\item \emph{Safety condition}: a play is safety winning if $occ(\rho) \cap F = \emptyset$, where $F\subseteq Q$ is the set of \emph{risk states}, and $occ(\rho)$ refers to the set of locations that appears in $\rho$.
\end{itemize}

\begin{figure}[t]
    \centering
     \includegraphics[width=0.7\columnwidth]{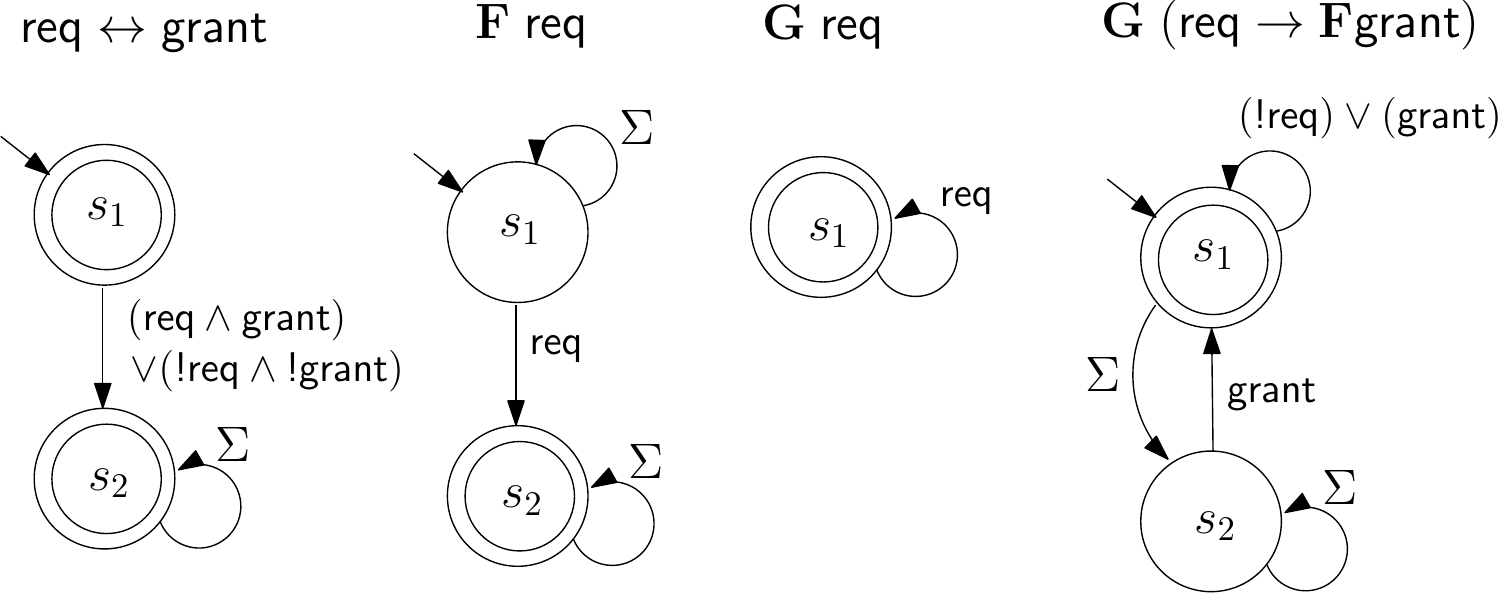}
      \caption{Some LTL formulas and their corresponding B\"{u}chi automaton, where $V_{in}=\{\textsf{req}\}$, $V_{out}=\{\textsf{grant}\}$, and $\Sigma= \{\langle \textsf{req,grant}\rangle,$$ \langle \textsf{!req,grant}\rangle, \langle \textsf{req,!grant}\rangle, \langle \textsf{!req,!grant}\rangle\}$.}
     \label{fig:Example.LTL}
\end{figure}



\subsection{LTL Synthesis}\label{sec.definition}

Given a set of input and output variables $V_{in}= \{\textsf{i}_1, \ldots, \textsf{i}_m\}$ and $V_{out}= \{\textsf{o}_1, \ldots, \textsf{o}_n\}$, together with an LTL formula $\phi$ on $V_{in}$ and $V_{out}$.  Let $\A_{\phi}$ be the corresponding B\"{u}chi automaton
of $\phi$. The LTL synthesis problem (where the environment takes the first move) asks the existence of a controller $f_{LTL}: (\mathbf{2}^{V_{in}}\times \mathbf{2}^{V_{out}})^{*}\times\mathbf{2}^{V_{in}} \rightarrow \mathbf{2}^{V_{out}}$ such that, for every input sequence $a= a_1a_2\ldots$, where $a_i \in \mathbf{2}^{V_{in}}$:
\begin{list1}
    \item Given the prefix $a_1$ produce $b_1$.
    \item Given the prefix $a_1 b_1 \ldots a_{k} b_{k} a_{k+1}$, produce $b_{k+1}$.
    \item The produced output sequence $b= b_1 b_2\ldots$ ensures that the word $\sigma = \sigma_1 \sigma_2 \ldots$, where $\sigma_i =  a_i b_i \in \mathbf{2}^{V_{in}\cup V_{out}}$, creates an accepting run of $\A_{\phi}$.
\end{list1}

Equivalently, we can adapt the same definition by applying it to UCW, i.e., given $\phi$, the controller should only allow the all runs created by produced word to visit the $\A_{\neg\phi}$ its final states only finitely often. Such a definition will be used later in bounded synthesis via safety games.

\noindent \textbf{(Example)} Consider the specification $\phi = \s{req} \leftrightarrow \s{grant}$. Based on our definition, any controller which outputs the same value as input in the first input-output cycle is considered to be a feasible solution.

\subsection{LTL Synthesis via UCW and Bounded-unroll Safety Games}

Given the LTL specification $\phi$, the roadmap of our method is to first construct the B\"uchi word automaton $\A_{\neg \phi}$, and view it as an UCW, i.e., if an $\omega$-word $\sigma$ can visit final states of $\A_{\neg \phi}$ only finitely often on every run, $\sigma$ is an $\omega$-word of $\phi$. Again assume that the construction of $\A_{\neg \phi}$ is provided.

\subsubsection{Creating Game-like Representations}

\begin{figure}[t]
    \centering
     \includegraphics[width=0.6\columnwidth]{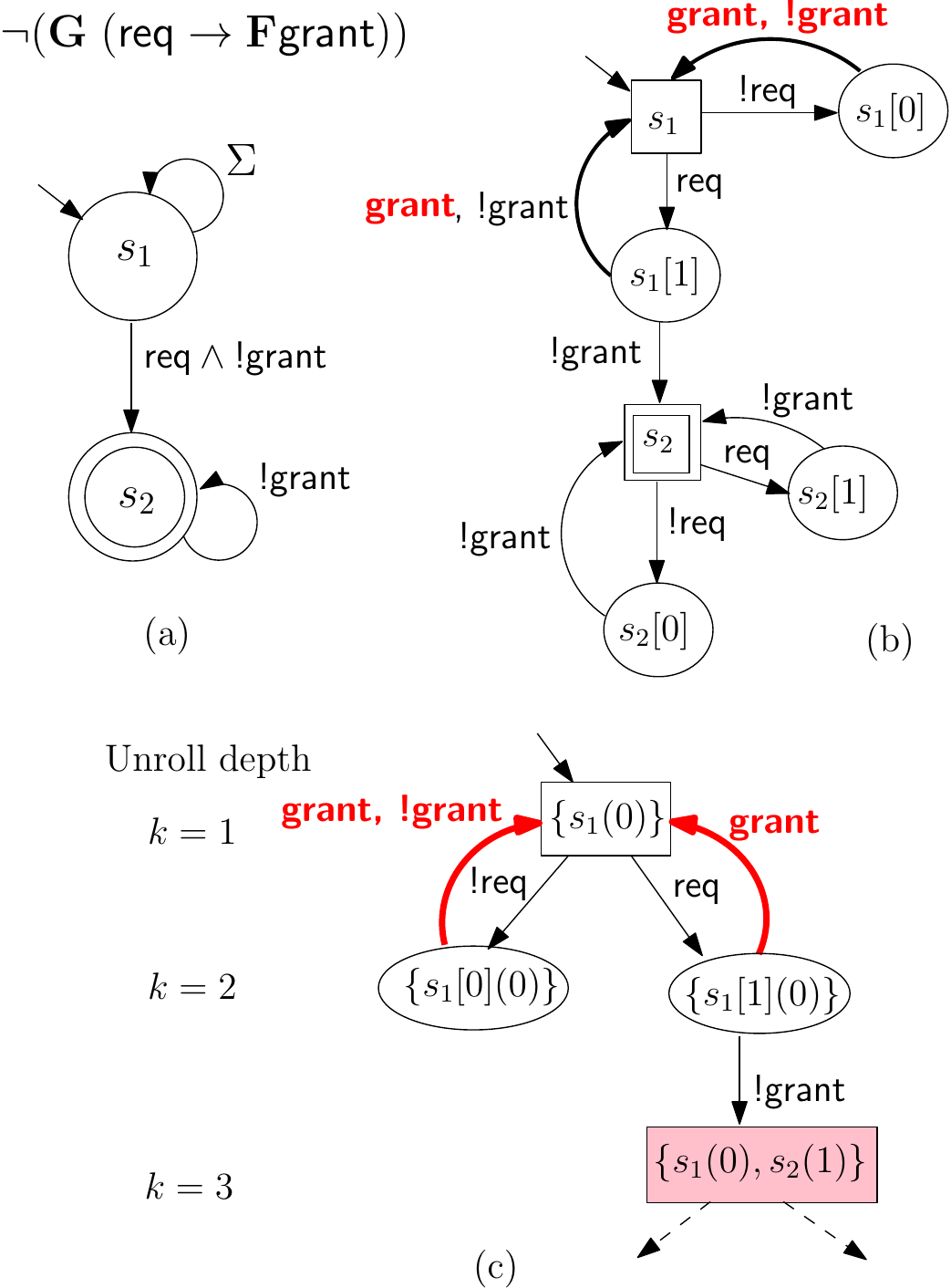}
      \caption{The B\"uchi automaton for $\neg\phi = \neg( \G (\s{req} \rightarrow \F \s{grant}))$ (a). The translated arena $\game$ (b) . The generated safety game (with equivalence class folding) using unroll of depth~$2$ (c), where the pink vertex is considered as a risk state.}
     \label{fig:Example.CoBuechi}
\end{figure}

\begin{algorithm}[t]
\begin{small}
\DontPrintSemicolon
\SetKwRepeat{doUntil}{do}{until}
\SetKwInOut{Input}{input}\SetKwInOut{Output}{output}
\Input{$\A_{\phi} = (Q_{\phi}, \Sigma= \Sigma_{in} \times \Sigma_{out}, q_{0_{\phi}}, \Delta_{\phi}, F_{\phi})$}
\Output{$\game_{\phi} = (Q, Q_0, E, q_0, \Sigma_{in}, \Sigma_{out}, L)$,  $F$ as accepting states}
\Begin{
     \texttt{let} $Q= Q_0 = E = L = \emptyset$\;
    /* Create environment locations in $\game$ */\;
     \nl$Q := Q \cup Q_{\phi}$\;
    /* Create control locations in $\game$ */\;
     \nl\For{$q \in Q_{\phi}$}{
        \For{$\s{in} \in \Sigma_{in}$}{
            $Q := Q \cup \{q[\s{in}]\}$\;
            $Q_0 := Q_0 \cup \{q[\s{in}]\}$\;
        }
    }
    /* Create deadlock state in  $\game$ */\;
    \nl$Q := Q \cup \{q_{dead}\}$\;
        \For{$\s{in} \in \Sigma_{in}$}{
            $Q := Q \cup \{q_{dead}[\s{in}]\}$\;
            $Q_0 := Q_0 \cup \{q_{dead}[\s{in}]\}$\;
        }
    /* Create environment edges and labeling in $\game$ */\;
    \nl\For{$q \in Q \setminus Q_0 $}{
      \For{$q[\s{in}] \in Q_0 $}{
            $E := E \cup (q, q[\s{in}])$\;
            $L(q, q[\s{in}]) := L(q, q[\s{in}]) \cup \{\s{in}\}$\;
      }
    }
    /* Create control edges and labeling in $\game$ */\;
     \nl\For{$(q_1, \langle \s{in}, \s{out}\rangle, q_2) \in \Delta_{\phi} $}{
        $E := E \cup (q_1[\s{in}], q_2)$\;
        $L(q_1[\s{in}], q_2) := L(q_1[\s{in}], q_2) \cup \{\s{out}\}$\;
    }
    /* Create missing control edges  $\game$ */\;
     \nl   \For{$q[\s{in}] \in Q_0 $}{
            \For{$\s{out} \in \Sigma_{out}$}{
            \If{$\not \exists q'\in Q\setminus Q_0$ such that $\s{out} \in L(q[\s{in}], q')$}{
            $E := E \cup (q[\s{in}], q_{dead})$\;
            $L(q[\s{in}], q_{dead}) := L(q_1[\s{in}], q_{dead}) \cup \{\s{out}\}$\;
            }
            }
        }
    \texttt{let} $q_0 = q_{0_{\phi}}$, and $\Sigma_{in}, \Sigma_{out}$ reuses contents in $\A_{\phi}$\;
    \texttt{return} ($\game_{\phi}$, $F_{\phi}$)\;
}
\caption{Translating an automaton into a game\label{algo.game.translation}}
\end{small}
\end{algorithm}

\begin{algorithm*}[t]
\begin{small}
\DontPrintSemicolon
\SetKwRepeat{doUntil}{do}{until}
\SetKwInOut{Input}{input}\SetKwInOut{Output}{output}
\Input{$\game = (Q, Q_0, E, q_0, \Sigma_{in}, \Sigma_{out}, L)$, $F: \{f_1, \ldots, f_n\}$, $k\in \mathbb{N}$}
\Output{$\game_{s}= (Q_s, Q_{s_0}, E_{s}, q_{s_0}, \Sigma_{in}, \Sigma_{out}, L_{s})$, $s_{risk}\in Q_s$}
\Begin{
     \texttt{let} $Q_s :=\emptyset$; $E_{s} := \emptyset$\;
    /* Create worklist with initial element */\;
    \texttt{let} $\s{worklist} := \emptyset$\;
    \nl \texttt{let} $q_{s_0}:= \{\s{update}(q_0(0, \ldots, 0), \s{list}(F))\}$\;
       \nl $\s{worklist.add}(q_{s_0},1)$\;
        $Q_s := Q_s \cup \{q_{s_0}, q_{risk}\}$\;
    \While{$list \neq \emptyset $}{
        \nl\texttt{let} $q=(\{q_1(v_{11}, \ldots, v_{1n}), \ldots, q_k(v_{k1}, \ldots, v_{kn})\},i) := \s{worklist.removeFirst}()$\;
        \If{$i \leq k$}{
            \If{$q \in Q \setminus Q_0$}{
                     /* create edges for environment vertex */\;
                     \For{$\s{in} \in \Sigma_{in}$}{
                            \nl \texttt{let}  $\s{succ} := \{q_1[\s{in}](v_{11}, \ldots, v_{1n}), \ldots, q_k[\s{in}](v_{k1}, \ldots, v_{kn})\}$\;
                            $Q_s := Q \cup \{\s{succ}\}$, $Q_{s_0} := Q_{s_0} \cup \{\s{succ}\}$;  $E_s := E_s \cup (q, \s{succ})$; $L(q, \s{succ}) := \{\s{in}\}$\;
                            \nl $\s{worklist.add}(\s{succ},i+1)$\;
                       }
                    } \Else{
                        /* create edges for control vertex*/\;
                        \For{$\s{out} \in \Sigma_{out}$}{
                            \texttt{let} \s{succ} := $\emptyset$\;
                            \For{$q_i(v_{i1}, \ldots, v_{in}) \in q$}{
                               \nl \For {$(q_i, q_{i}') \in E$ where $\s{out} \in L(q_i, q_{i}')$}{
                               \nl     $\s{succ} := \s{succ} \uplus \s{update}(q_{i}'(v_{i1}, \ldots, v_{in}),  \s{list}(F))\}$\;
                               }
                            }
                             /* Check if a vertex of same equivalence class has appeared */\;
                            \nl\If {$Q_s.\s{getEquiv}(\s{succ}) \neq \s{null}$}{
                             $E_s := E_s \cup (q, Q_s.\s{getEquiv}(\s{succ}))$; $L(q, Q_s.\s{getEquiv}(\s{succ})) := \{\s{out}\}$\;
                            } \Else{
                                $E_s := E_s \cup (q, \s{succ})$; $L(q, \s{succ}) := \{\s{out}\}$\;
                                $Q_s := Q_s \cup \{\s{succ}\}$\;
                               \nl $\s{worklist.add}(\s{succ},i+1)$\;
                            }

                        }
                    }
            } \Else {
          \nl      $E_s := E_s \cup (q(v_1, \ldots, v_n), q_{risk})$\; 
            }
    }
    \texttt{return} $(Q_s, Q_{s_0}, E_{s}, q_{s_0}, \Sigma_{in}, \Sigma_{out}, L_{s}), q_{risk}$\;
}
\caption{Depth-$k$ unroll to safety games\label{algo.game.unroll}}
\end{small}
\end{algorithm*}

The first step, after creating $\A_{\neg \phi}$, is to apply Algorithm~\ref{algo.game.translation} to create a game representation $\game_{\neg \phi}$. Consider again the synthesis process of  $\phi = \G (\s{req} \rightarrow \F \s{grant})$. Figure~\ref{fig:Example.CoBuechi}~(a) and~(b) show the corresponding $\A_{\neg \phi}$ and the translated game. A controller will, for every input sequence, produce the corresponding output sequence to ensure that all paths in the generated game will visit final states in $\game_{\neg \phi}$ only finitely often. Not difficultly, we can observe a solution highlighted at Figure~\ref{fig:Example.CoBuechi}~(b), which outputs $\s{grant}$ at $s_1[1]$ (i.e., when input equals $\s{req}$) and outputs $\s{grant}$ or $\s{!grant}$ at $s_1[0]$ (i.e., when input equals $\s{!req}$). Outputting  $\s{grant}$ at $s_1[1]$ ensures that $s_2$ is never visited from the initial state.

\subsubsection{Bounded Unroll into Safety Games}

Once when the game representation $\game_{\neg \phi}$ of $\A_{\neg \phi}$ is generated, we then perform a $k$-bounded unroll by exploring all finite words of size $k/2$ defined in $\A_{\neg \phi}$ to create a safety game. For all vertices whose descendants are not fully expanded, we consider them to be risk states.  We define \emph{equivalence relation} which enables to treat the two different vertices as the same one. By doing so, the unfolding in general does not create an arena with a tree structure but rather an arena with loops.

Given $\game_{\neg \phi} = (Q, Q_0, E, q_0, \Sigma_{in}, \Sigma_{out}, L)$ with final states $F_{\neg \phi}= \{f_1, \ldots, f_n\}$ provided by Algorithm~\ref{algo.game.translation}, a node (i.e., an equivalence class) in the translated safety game uses a set $\{q(v_1, \ldots, v_n)\}$ as its identifier, where $q\in Q$ is the last visited vertex and $v_i \in \mathbb{N}_0$ is the number of visited times for final state $f_i$, for a finite word whose corresponding state sequence ends at $q$. Algorithm~\ref{algo.game.unroll} shows a worklist algorithm which performs $k$-bounded unroll to create a new safety game.

The worklist algorithm first pushes the initial state into the list (line~1,2). Then it continuously removes elements in the list to perform the expansion. As each state is associated with a number when it is placed in the worklist (line~2,5,9) and will not produce new elements to the list when the value exceeds $k$, the algorithm guarantees termination. If the element taken from the list is an environment vertex, then new vertices are created based on all possible input values (line~4). Otherwise, the algorithm considers, given an output valuation $\s{out}$, what are all possible successor states (line~6,~7). In line~7, for the ease of understanding, we use the following functions.
\begin{list1}
\item $\s{update}(q(v_0, \ldots, v_n), \s{list}(F))$ updates $v_i$ by $v_i+1$, if in the list representation of $F$, the index of $q$ equals $i$. It is also used in line~1, i.e., if the initial state is a final state, then the value should start with $1$ rather than~$0$.
\item Given $q=\{q_1(v_{11}, \ldots, v_{1n}), \ldots,  q_k(v_{k1}, \ldots, v_{kn})\}$, the operation $q \uplus q_i(v_{i1}', \ldots, v_{in}')$ replaces $q_i(v_{i1}, \ldots, v_{in})$ in $q$ by $q_i(\s{max}(v_{i1},v_{i1}'), \ldots, \s{max}(v_{in}, v_{in}'))$. For example $\{q_1(1,0), q_2(2,3)\} \uplus q_1(0,1)$ we derive $\{q_1(1,1), q_2(2,3)\}$. Originally, having two elements $q_1(1,0)$ and $q_1(0,1)$ means that to visit $v_1$, there exists two runs where the first traverses the first final state and the other traverses the second final state. We do not maintain two separate information but rather use an overapproximation. It is the reason why our method is called \emph{merge-over-all-paths}, as such a technique is borrowed from static analysis in program verification.
\end{list1}

Therefore, in line~7, when the state is updated from $q_i$ to $q_{i}'$, at first the algorithm produces $q_{i}'(v_{i1}, \ldots, v_{in})$. Then it uses function $\s{update}()$ to update the number of visited final states considering $q_{i}'$. Then it uses $\uplus$ to merge the result of newly computed $q_{i}'$ to existing results. Line~8 checks if there exists a vertex in $Q_s$ which has the same value of $\s{succ}$ (using $Q_s.\s{getEquiv}(\s{succ})$). If such a vertex exists, then connect all outgoing edges to the existed vertex rather than $\s{succ}$.

\textbf{(Example)} Consider the following specification: $\phi_{TG}: \G (\s{req} \rightarrow ( \s{grant} \vee \X \s{grant})) \wedge \G( \s{grant} \rightarrow \X \s{!grant})$, which specifies that when a request arrives, a grant should be issued $\s{true}$ immediately or with at most one unit delay. However, it is disallowed to issue to consecutive grants. Figure~\ref{fig:Example.MergeAllPath} shows the B\"uchi automaton of the complement specification $\neg\phi_{TG}$. Two paths (1) $s_1\rightarrow s_2\rightarrow s_4$ and (2) $s_1\rightarrow s_3\rightarrow s_4$ reflect the erroneous scenario where in (1) a request is not granted in time, and in (2) there are two consecutive grants.
Let $\s{list}(F) = (s_2, s_3, s_4)$. Figure~\ref{fig:Example.MergeAllPathSafetyGame} shows the result of unrolled safety game.
$\game_{\neg \phi_{TG}}$ starts with initial state $s_1$, and it has not visited any final state. Therefore the algorithm creates
vertex $\{s_1(0,0,0)\}$ and add to the worklist (line~1). Then the algorithm creates $\{s_1[0](0,0,0)\}$ and $\{s_1[1](0,0,0)\}$, reflecting different input values (line~4). $s_1$ can move to $\{s_1, s_3\}$ via $\s{grant}$ when receiving input $\s{req}$. Therefore, we link $\{s_1[1](0,0,0)\}$ to a new environment vertex $\{s_1(0,0,0), s_3(0,1,0)\}$, where $s_3(0,1,0)$ denotes that final state $s_3$ has been reached once (recall that $s_3$ is the second element in $\s{list}(F)$, so $s_3(0,1,0)$ changes to $s_3(0,1,0)$ after \s{update}()). For $s_1$ and $s_3$, given any input, if output equals $\s{!grant}$, $s_3$ cannot proceed while $s_1$ can move to $s_1$ or $s_2$. Therefore, by responding $\s{!grant}$,  $\{s_1[0](0,0,0), s_3[0](0,1,0)\}$ and $\{s_1[1](0,0,0), s_3[1](0,1,0)\}$ will move to $\{s_1(0,0,0), s_2(1,0,0)\}$. In this example, an unroll of depth $5$ is sufficient to create a controller.

\begin{figure}[t]
    \centering
     \includegraphics[width=0.75\columnwidth]{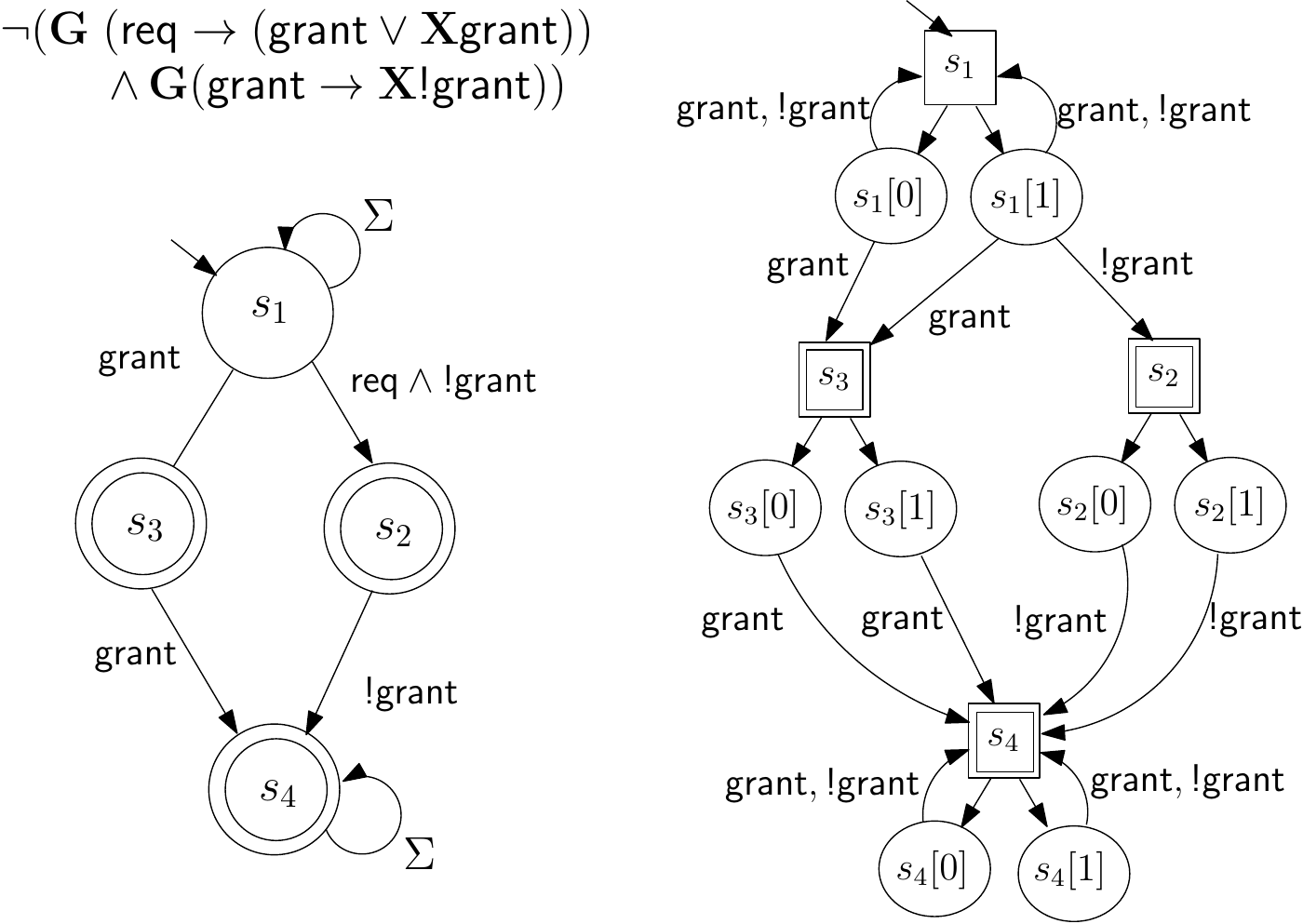}
      \caption{The B\"uchi automaton $\A_{\neg \phi_{TG}}$ for $\neg \phi_{TG}$ (left) and the translated arena $\game_{\neg\phi_{TG}}$ (right).}
     \label{fig:Example.MergeAllPath}
\end{figure}

\begin{figure}
    \centering
     \includegraphics[width=0.75\columnwidth]{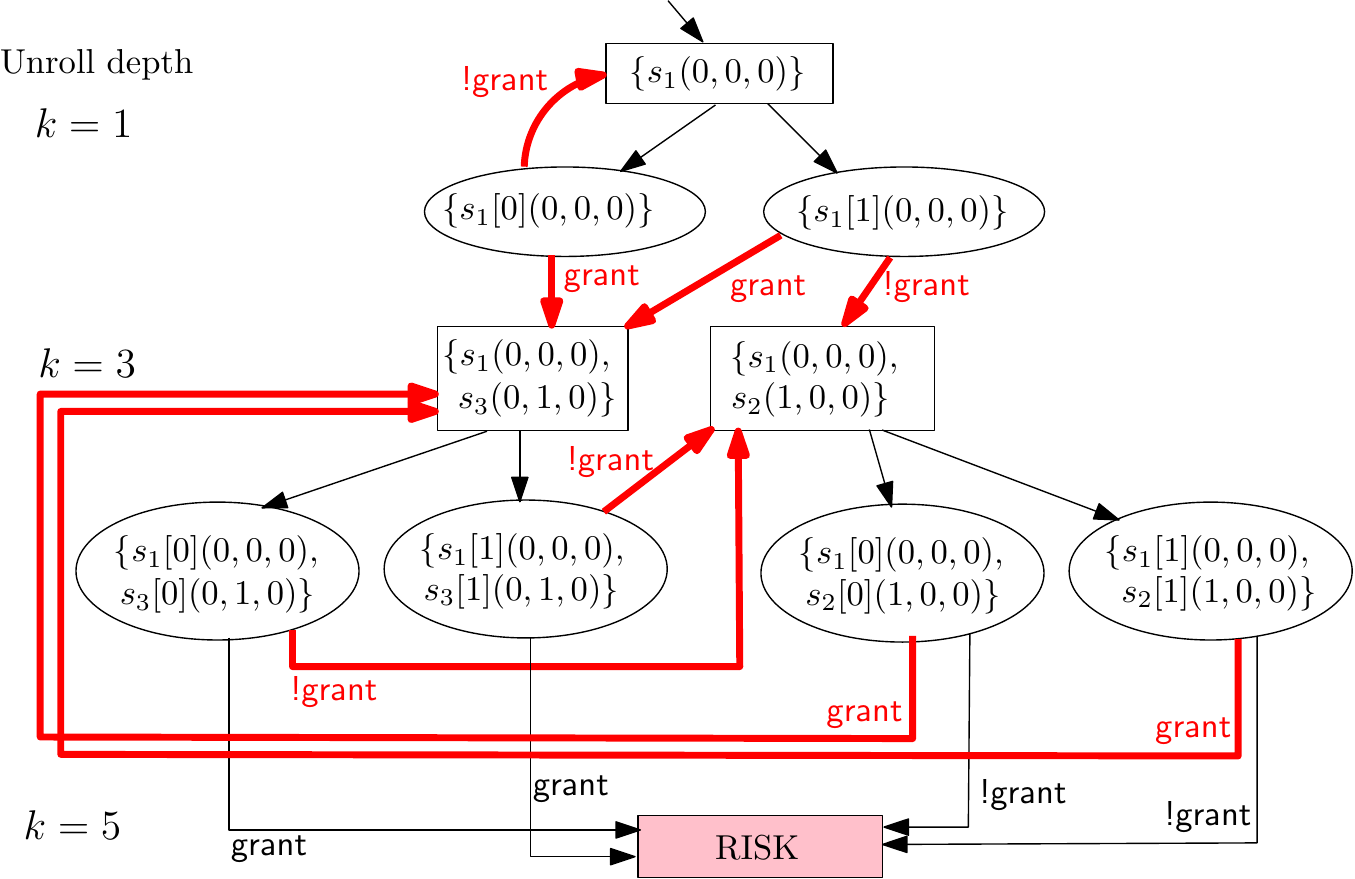}
      \caption{The generated safety game (with equivalence class folding) from $\game_{\neg\phi_{TG}}$ using unroll of depth~$3$.}
     \label{fig:Example.MergeAllPathSafetyGame}
\end{figure}

\end{document}